\documentclass{elsart}
 \usepackage{amsmath}
  \usepackage{amsfonts}
\usepackage{amssymb}
 \usepackage{graphicx}
  
 \newcommand{\singlefig}{.75\textwidth}
 
  \newcommand{\ii}{\mathrm{i}}

\begin{document}
\begin{frontmatter}
\journal{Physics Letters A}
\date{9 September, 2003}
\title{Nonlinear charge transport in DNA mediated by twist modes}
\author[GFNL]{F Palmero},
\author[GFNL]{JFR Archilla \thanksref{archilla}}, 
\author[Berlin]{D Hennig} and 
\author[GFNL]{FR Romero}
\address[GFNL]{Nonlinear Physics Group (GFNL) of the University of Sevilla, Spain}
\address[Berlin]{Freie Universit\"at Berlin, \\ Fachbereich Physik, Arnimallee 14,
14195-Berlin, Germany}
\thanks[archilla]{Corresponding author. 
Departamento de F\'{\i}sica Aplicada I. ETS Ingenier\'{\i}a
Inform\'atica. Avda Reina Mercedes
s/n, 41012-Sevilla, Spain. 
 Email: archilla@us.es}


\begin{abstract}
Recent works on localized charge transport along DNA, based on a
three--dimensional, tight--binding model (Eur. Phys. J. B 30:211, 2002;
Phys. D 180:256, 2003), suggest that charge
transport is mediated by the coupling of the radial and electron
variables. However, these works are based on a linear
approximation of the distances among nucleotides, which forces for
consistency the assumption that the parameter $\alpha$, that
describes the coupling between the transfer integral and the
distance between nucleotides is fairly small. In the present letter we
perform an improvement of the model which  allows larger values of
$\alpha$, showing that there exist two very different regimes.
Particulary, for large $\alpha$, the conclusions of the previous
works are reversed and  charge transport can be produced only by
the coupling with the twisting modes.
\end{abstract}

\end{frontmatter}

\section{Introduction}

Charge transport in DNA is an interesting subject because of its
role for  biological functions such as DNA repair after radiation
damage and biosynthesis, and, on the other hand, because of the
possibility of building electronic devices based on
biomaterials~\cite{Ratner99,Fink99,Tran00,Braun98,Porath00}. When
a charge moves along the double DNA strand it is accompanied by a
local deformation of the molecule, forming a polaron. Simple, but
powerful models of DNA are based on the one proposed by
Peyrard--Bishop \cite{PB89}, in which the only variables are the
distances  between bases within each base pair. The corresponding
hydrogen bonds are described by Morse potentials. In the context
of polaron dynamics, a tight--binding system coupled with the bond
distances is used for the description of the charge dynamics. In
this case, as the deformation of the helix is small a harmonic
approximation of the bond potentials is considered enough giving
rise to a Holstein model~\cite{Holstein59,KAT98}.

However, the helical structure of DNA  is thought to play a key
role in its functional processes and recently steric models of the
molecule have been introduced in the context of vibrational
motion~\cite{Barbi98,Barbi99a,Barbi99b,Cocco99,Cocco00} and charge
transport~\cite{H02,HAA03,AHA03}. As it will be shown in this article,
the value of the parameter $\alpha$ that couples the electronic
transfer integral and the distance between nucleotides is of
particular importance, determining to a great extent the degree of
charge localization and its mobility properties. However there is
no experimental evidence of its value. Previous
works~\cite{H02,HAA03,AHA03} have dealt with very small values of
$\alpha$, because the angular deformations of the helix had to be
small in compliance with the linear approximations performed in
the expressions of the distances between nucleotides. One
conclusion of these works are that charge transport in DNA is
mediated by coupling of the charge carrying unit with the
hydrogen--bond deformations, and that this transport survives to a
small amount of parametric and structural disorder. However, we
show in this work, taking into account the full equations of the
model, that for larger values of $\alpha$ a completely different
regime appears. In this regime charge transport is mediated  by
local winding and unwinding of the helix. Moreover, this regime is
more robust with respect to disorder than the one mediated  by the
coupling with the hydrogen--bond distances.
\section{The model}
In order to describe the dynamic of a DNA chain, we consider a
variant of the {\em twist-opening}
model~\cite{Barbi98,Barbi99a,Barbi99b}, which itself is a
modification of the Peyrard--Bishop model taking into account the
helical structure of the molecule and the torsional deformations
induced by the opening of the base pairs. Each group of sugar,
nucleotide and base will be considered as a single,
non--deformable object, with mass $m$, which is an averaged
estimation of the nucleotide mass. As we are interested in base
pair vibrations and not in acoustic motions, we fix the center of
mass of each base pair, $i.e.$, the two bases in a base pair are
constrained to move symmetrically with respect to the molecule
axis. Moreover, the distances between two neighboring base pair
planes shall be treated as fixed, because in the axial direction
DNA seems less deformable than within the base pair
planes~\cite{Barbi98}.

\begin{figure}
  \begin{center}
    \includegraphics[width=\singlefig]{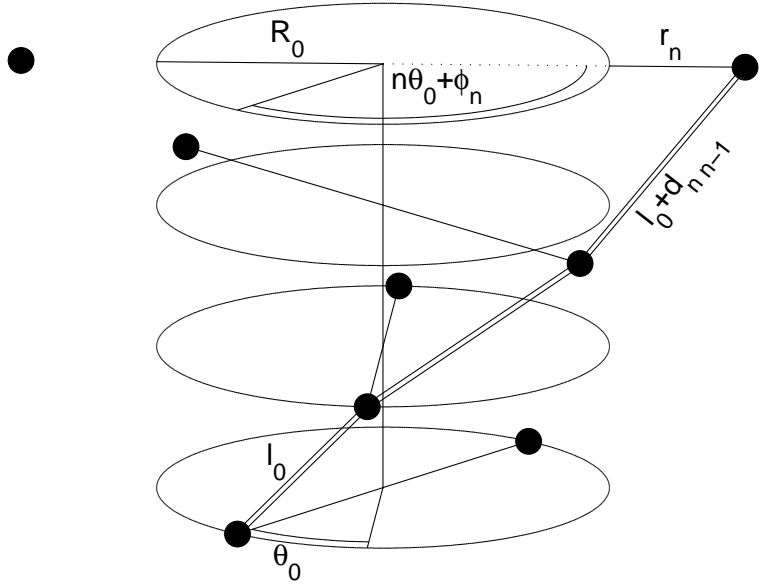}
  \end{center}
  \caption{Sketch of the model. Filled circles  represent  bases. The variables
  used in the text are displayed.}
  \label{model}
\end{figure}

In our model, all bond potentials will be treated as harmonic
ones. This can be justified because charge transport is related
with small deformations of the double chain. Furthermore, as the
angular twist and the radial vibrational motion evolve on two
different time scales, they can be considered as decoupled degrees
of freedom in the harmonic approximation~\cite{Cocco00}. Thus, the
position of the n--th base pair is represented by the variables
$(r_n,\phi_n)$, where $r_n$ represents the radial displacement of
the base pair from the equilibrium value $R_0$, and $\phi_n$ its
angular displacement from equilibrium  angles with respect to a
fixed external reference frame. A sketch of the model is shown in
Fig.~\ref{model}.

 The Hamiltonian of the system is given by $
\widehat{H}=\widehat{H}_{el}+\widehat{H}_{rad}+\widehat{H}_{twist}$,
where $\widehat{H}_{el}$ corresponds to the part related with the
charged particle described by a tight--binding system:
$\widehat{H}_{el}=\sum_nE_n|n \rangle \langle n|-V_{n-1,n}|n-1
\rangle \langle n|-V_{n+1,n}|n+1 \rangle \langle n|$,
where $|n \rangle $ represents a localized state of  the charge
carrier at the $n^{th}$ base pair. The quantities $\{V_{n,n-1}\}$
are the nearest--neighbor transfer integrals along  base pairs,
and $\{E_n\}$ are the energy on--site matrix elements. A general
electronic state is given by $|\Psi\rangle=\sum_n
c_n(t)|n\rangle$, where $c_n(t)$ is the probability amplitude of
finding the charged particle in the state $|n\rangle$. The time
evolution of the $\{c_n(t)\}$ is obtained from the
Schr$\ddot{o}$dinger equation
$\ii \hbar(\partial |\Psi\rangle/ \partial
t)=\widehat{H}_{el}|\Psi\rangle$

The nucleotides are large molecules and their molecular motions
are slow compared to the one of a charged particle, the lattice
oscillators may be treated classically and $\widehat{H}_{rad}$ and
$\widehat{H}_{twist}$, describing the radial and torsional
contributions to $\widehat{H}$,  are, {\em de facto}, classical
Hamiltonians. They are  given by (omitting the hat on them):
\begin{eqnarray}
H_{rad} &=& \sum_n \left[ \frac{1}{2M} (p^r_n)^2+\frac{M
\Omega_r^2}{2}r_n^2 \right],\\  H_{twist} &=& \sum_n  \left[
\frac{1}{2J} (p^{\phi}_n)^2+\frac{J \Omega_\phi^2}{2}\
(\phi_n-\phi_{n-1})^2 \right],
\end{eqnarray}
where $p^r_n$ and $p^{\phi}_n$ are the conjugate momenta of the
radial and angular coordinates, respectively.  $M=2m$ and $J=M\,R_0^2$, are the
mass and the inertia moment of each base pair respectively,
$\Omega_r$ is the linear radial frequency, and $\Omega_\phi$ is
the linear twist frequency. We represent by
$\theta_{n,n-1}=(\phi_n-\phi_{n-1})$, the deviation of the
relative angle between two adjacent base pairs from its
equilibrium value $\theta_0$.

The interaction between the electronic variables and the structure
variables $r_n$ and $\phi_n$ arises from the dependence of the
matrix elements $E_n$ and $V_{n,n-1}$ on them. The first ones are
given by~\cite{KAT98} $E_n = E_n^0+k r_n$, expressing the
variation of the on--site electronic energies $\{E_n^0\}$ with the
radial deformations. We assume that the transfer matrix elements become smaller
when the distances between nucleotides grows. That is,
$V_{n,n-1}$ depends on the distances between two consecutive bases
along a strand as
 $V_{n,n-1} = V_0(1-\alpha d_{n,n-1})$, with $d_{n,n-1}$ given by
\begin{eqnarray}
\label{distance}
\nonumber d_{n,n-1} & = & [a^2+(R_0+r_n)^2+ (R_0+r_{n-1})^2-
\\
& & 2(R_0+r_n)(R_0+r_{n-1})
\cos(\theta_0+\theta_{n,n-1})]^{1/2}-l_0,
\end{eqnarray}
with
 $
l_0=(a^2+4R_0^2 \sin^2(\theta_0/2))^{1/2},
 $
and $a$ is the vertical distance between base pairs. The parameter
$\alpha$ describes the influence of the distances between
nucleotides on the transfer integrals. In this paper we do not perform
the linear
approximation of the distances $d_{n,n-1}$ considered in
Refs.~\cite{HAA03,AHA03}, allowing for
larger values of the perturbations.

Realistic parameters for DNA are given in
Refs.~\cite{Barbi99a,Stryer95}.
 We have considered: $a= 3.4 \AA$, $m=300$ amu, $R_0= 10
\AA$, $\Omega_r=8 \times 10^{12}$ s$^{-1}$, $\Omega_{\phi}=9 \times
10^{11}$ s$^{-1}$, and $V_0=0.1$ eV. We consider
 $\alpha$ and $k$ as adjustable parameters.

The present model is a modification of a previous one proposed in
Refs.~\cite{H02,HAA03,AHA03}, in order to study the transport of
charge by polarons in DNA. In these works, only changes on the
relative angular velocity between two neighboring base pairs,
$\dot{\theta_{n,n-1}}$, are taken into account in the rotational
kinetic energy, ignoring the contribution of the rest of the
terms. Also, in the cited references, only very small
displacements around the equilibrium position of the base pairs
are considered, and the distances $d_{n,n-1}$ can be approximate
by their Taylor expansion up to first order. In this paper we
allow the spatial variables to have (relatively) large values
and therefore we use the full nonlinear Eqs.~(\ref{distance}) for
the distances $d_{n,n-1}$.

We scale the time according to $t \rightarrow \Omega_r t$, and
introduce the dimensionless quantities:
 $\tilde{r}_{n}=r_n\,(M \Omega_{r}^{2}/V_0)^{1/2}$,
 $\tilde{k}_{n}=k_n/(M\Omega_{r}^{2}V_0)^{1/2}$,
 $ \tilde{E_n}=E_n/V_0$,
 $\tilde{\Omega}=\Omega_{\phi}/\Omega_r$,
 $\tilde{V}=V_0/(J\,\Omega_{r}^2)$,
 $\tilde{\alpha}=\alpha\, (V_0/M\,\Omega_r^2)^{1/2}$,
 $\tilde{R}_{0}=R_0\,(M\,\Omega_r^2/V_0)^{1/2}$.
The scaled dynamical equations of the system,  from which we have
omitted the tildes, are:
\begin{eqnarray}
\lefteqn{\ii\,\tau\dot{c}_{n}=(E_n\,+k\,r_n)\,c_n}\nonumber\\
\label{electron} & &-(1-\alpha\,d_{n+1,n})\,c_{n+1}
-(1-\alpha\,d_{n\,n-1})\,c_{n-1}\,,
\\  \nonumber \lefteqn{\ddot{r}_{n}=-r_n-k\,|c_n|^2} \\
& & -\, \alpha\, \left[\frac{\partial d_{n,n-1}}{\partial
r_n}(c_{n}^*c_{n-1}+c_{n}c_{n-1}^*)+\frac{\partial
d_{n+1,n}}{\partial r_n}(c_{n+1}^*c_{n}+c_{n+1}c_{n}^*) \right]\,,
\label{radial} \\
\nonumber \lefteqn{\ddot{\phi}_n
=-\Omega^2\,(2\phi_n-\phi_{n-1}-\phi_{n+1})}
\\  & & -\,
  \alpha\,V\,\left[\frac{\partial d_{n,n-1}}{\partial \phi_n}
(c_{n}^*c_{n-1}+c_{n}c_{n-1}^*)+ \frac{\partial
d_{n+1,n}}{\partial \phi_n}(c_{n+1}^*c_{n}+c_{n+1}c_{n}^*)\right]\,,
\label{angular}
\end{eqnarray}
and the quantity $\tau=\hbar\,\Omega_{r}/V_0$ determines the time
scale separation between the fast electron motion and the slow
bond vibrations. In the limit case of $\alpha=0$ and uniform
$E_n=E_0$ the set of coupled equations represents the Holstein
system, widely used in studies of polaron dynamics in
one-dimensional lattices, and for $\alpha=k=0$, and random $E_n$,
the Anderson model is obtained. Using the expectation value for
the electronic contribution to the Hamiltonian, the new
Hamiltonian $\overline{H}=\langle \phi |\widehat{H}|\phi\rangle
/V_0$ is given by
\begin{eqnarray}
\nonumber \overline{H}&=&\sum_{n}\Big\{
\frac{1}{2}(\dot{r}_n^2+r_n^2)+\frac{R_0^2}{2}[\dot{\phi_n^2}+\Omega^2(\phi_n-\phi_{n-1})^2]+
\\
 & &  (E_n^0+k r_n)|c_n|^2-(1-\alpha d_{n,n-1})(c_{n}^*c_{n-1}+c_{n}c_{n-1}^*) \Big \}.
\end{eqnarray}
The values of the scaled parameters are $\tau=0.053$,
$\Omega^2=0.013$, $V=2.5 \times 10^{-4}$, $R_0=63.1$ and
$l_0=44.5$. We fix the value  $k=1$ and consider the parameter
$\alpha$ as adjustable, its value complying  with the hypothesis
of small deformations of the helix.

\section{Stationary polaron-like states}
As a first step in our  study of  nonlinear charge transport in
this DNA model, we focus our interest in localized stationary
solutions of Eqs.~(\ref{electron}-\ref{angular}). Since the
adiabaticity parameter $\tau$ is small, the fastest variable are
the $\{c_n\}$, with a characteristic frequency (the linear
frequency of the uncoupled system) of order $1/\tau \sim 19$,
followed by the $\{r_n\}$ with frequency unity, and the
$\{\phi_n\}$ with $\Omega_{\phi}\sim 0.11$. Using the
Born--Oppenheimer approximation, we can suppose initially that
$r_n$ and $\phi_n$ are constant in order to obtain the stationary
localized solutions. For this purpose, we use a modification of
the numerical method outlined in Refs.~\cite{KAT98,VT01}. We
substitute in Eq.~(\ref{electron}) $c_n=\Phi_n
\exp(-\ii\,E\,t/\tau)$, with time--independent $\Phi_n$'s, and we
obtain a nonlinear difference system $E \Phi=\widehat{A}\Phi$,
with $\Phi=(\Phi_1,...,\Phi_N)$, from which a map
$\Phi'=\widehat{A}\Phi/\|\widehat{A}\Phi\|$ is constructed,
$\|.\|$ being the quadratic norm.

Thus, using Eqs.~(\ref{electron}-\ref{angular}), the stationary
solutions must be  attractors of the  map:
\begin{eqnarray}
\lefteqn{ r'_{n}=-k\,|c_n|^2} \nonumber \\
& & -\, \alpha\, \left[\frac{\partial d_{n,n-1}}{\partial
r_n}(c_{n}^*c_{n-1}+c_{n}c_{n-1}^*)+\frac{\partial
d_{n+1,n}}{\partial r_n}(c_{n+1}^*c_{n}+c_{n+1}c_{n}^*) \right]\,,
\label {radialmap} \\ \nonumber
\lefteqn{\phi'_n=\frac{1}{2}(\phi_{n+1}+\phi_{n-1})}
\\  & & -\,
  \frac{\alpha V}{2 \Omega^2}\,\left[\frac{\partial d_{n,n-1}}{\partial \phi_n}
(c_{n}^*c_{n-1}+c_{n}c_{n-1}^*)+ \frac{\partial
d_{n+1,n}}{\partial \phi_n}(c_{n+1}^*c_{n}+c_{n+1}c_{n}^*)\right]\,,
\label
{angularmap} \\
\lefteqn{ c'_{n}=\frac{[(E_n\,+k\,r'_n)\,c_n
\label{electronmap} -(1-\alpha\,d'_{n+1,n})\,c_{n+1}
-(1-\alpha\,d'_{n\,n-1})\,c_{n-1}]}{\|\{(E_n\,+k\,r'_n)\,c_n
-(1-\alpha\,d'_{n+1,n})\,c_{n+1}
-(1-\alpha\,d'_{n\,n-1})\,c_{n-1}\}\|}\,,}
\end{eqnarray}
where $d'=d(r',\phi')$. In order to achieve convergence, it has been necessary to use
the result of
an interaction in the geometrical variables as a seed in the map corresponding
to the probability amplitude. The starting point is a completely
localized state given by $c_n=\delta_{n,0}$, $r_n=0$ and
$\phi_n=0$, $\forall n$. The map is applied until convergence is
achieved. In this way both stationary solutions and their energies
$E$ are obtained.

We analyze the ordered case, i.e., $E_n^0=E_0$, that arises in
synthetic DNA (the constant $E_0$ can be set to zero by means of a
gauge transformation), and the disordered case. In the latter, we
introduce diagonal disorder in the on--site electronic energy
$E_n^0$ by means of a random potential $E_n^0\in [-\Delta E,\Delta
E]$, with mean value zero and different interval sizes $\Delta E$. Such
deviations can be caused by inherent disorder in the surrounding water and by the
inhomogeneous distribution of counterions along the DNA duplex \cite{BC01}.

In the ordered case, we have found that in the ground state the
charge is fairly localized at one site, and the amplitudes decay
monotonically and exponentially with growing distance from the
central site. The associated patterns of the static radial and
relative angular displacements are similar. In the disordered
case, the localized excitation patterns do not change
qualitatively. However, as the translational invariance is broken
by the disorder, the localized excitation pattern is not symmetric
with respect to a lattice site, in opposition to the ordered case.
\begin{figure}
  \begin{center}
  \begin{tabular}{cc}
  \includegraphics[width=0.8\textwidth, height=0.3\textheight]{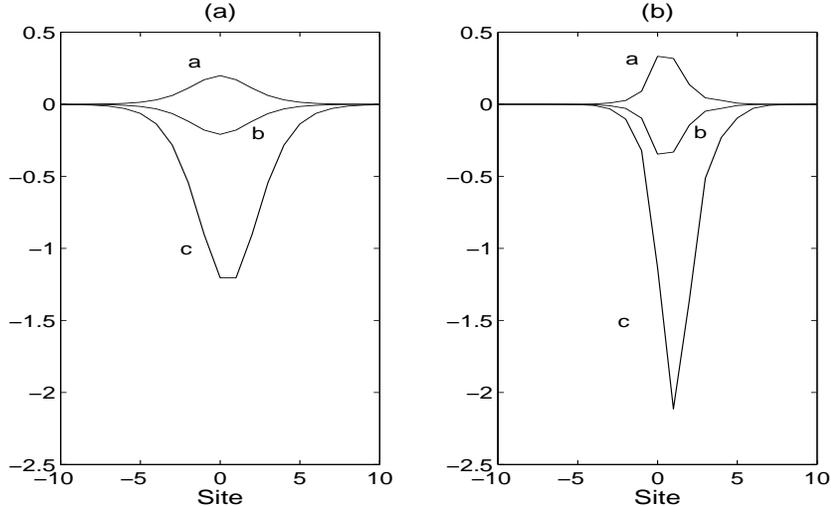}
    \end{tabular}
  \end{center}
  \caption{Profiles of the stationary state for $\alpha=0.05$. (a) Ordered case. (b)
Disordered case for $\Delta E=0.01$ eV.
  a) Electronic probabilities $|c_n|^2$. b) Radial displacements
  $r_n$, c) Arc displacements $R_0\,\theta_{n,n-1}$. The displacements are in scaled units
  equivalent to
  $\simeq 0.16 \AA$.}
  \label{stationary}
\end{figure}

In order to study the localization of the excitation, we introduce
the degree of localization $L$ of an amplitude pattern $\{u_n\}$
as
\begin{equation}
L=\frac{\sum_n|u_n|^2}{(\sum_n|u_n|)^2},
\end{equation}
and the participation number $P$ as its inverse $P=1/L$. This
magnitude gives an estimate of the number of excited oscillators
in the chain.

\begin{figure}[h]
\begin{center}
\includegraphics[width=\singlefig]{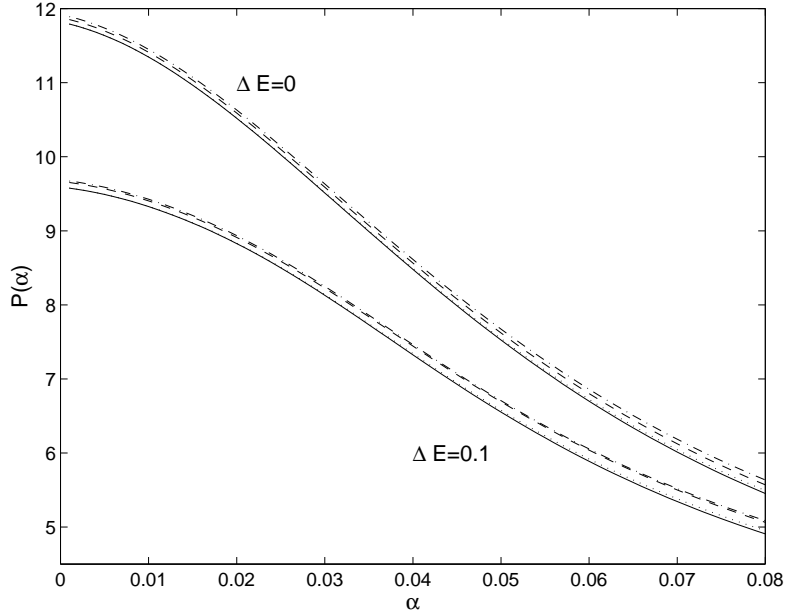}
\end{center}
\caption{Participation number $P$ as a function of $\alpha$ for
the electronic probabilities $|\Phi_n|^2$ (solid line), radial
displacements $r_n$ (dotted line), angular displacements $\theta_{n,n-1}$
(dashed line), and density of energy (dashdot line) of the ground
state for the ordered case ($\Delta E=0$) and the
disordered one ($\Delta E=0.01$ eV)} \label{local}
\end{figure}

We have studied the participation numbers of the electronic
probabilities $|\Phi_n|^2$, the radial displacements $r_n$, the
angular displacements $\theta_{n,n-1}$, and the density of energy as functions
of the parameter $\alpha$. As Fig.~\ref{local} shows, for each
given value of  $\alpha$ and the disorder amplitude, the
participation numbers of all these magnitudes are very close. When
$\alpha$ increases, the participation decreases and therefore the
localization increases. Comparing the ordered and the disordered
case, we can observe that the localization is enhanced with the
disorder, due to Anderson localization~\cite{Anderson58}.

\section{Charge transport in the absence of disorder}
In this section, we study charge transport  along the double
strand by moving polarons. Once a stationary state is obtained, it
can be moved under certain conditions. There exists a systematic
way to do it, known as the {\em pinning mode}~\cite{CAT96}, which
consists of perturbing the (zero) velocities of the ground state
with localized, spatially antisymmetric modes obtained in the
vicinity of a bifurcation. This method leads to moving entities
with very low radiation but has the inconvenience of being
applicable only in the neighborhood of certain values of the
parameters. Instead, we use the discrete gradient
method~\cite{IST02}, perturbing  the (zero) velocities of the
stationary state $\{\dot{r}_n(0)\}$, $\{\dot{\phi}_n(0)\}$ in a
direction parallel to the vectors $(\nabla r)_n=(r_{n+1}-r_{n-1})$
and/or $(\nabla \phi)_n=(\phi_{n+1}-\phi_{n-1})$. Although this
method does not guarantee  mobility, it nevertheless proves to be
successful in a wide parameter range.
\subsection{Radial movability regime}
\begin{figure}
  \begin{center}
 \includegraphics[height=0.85\textheight,width=\textwidth]{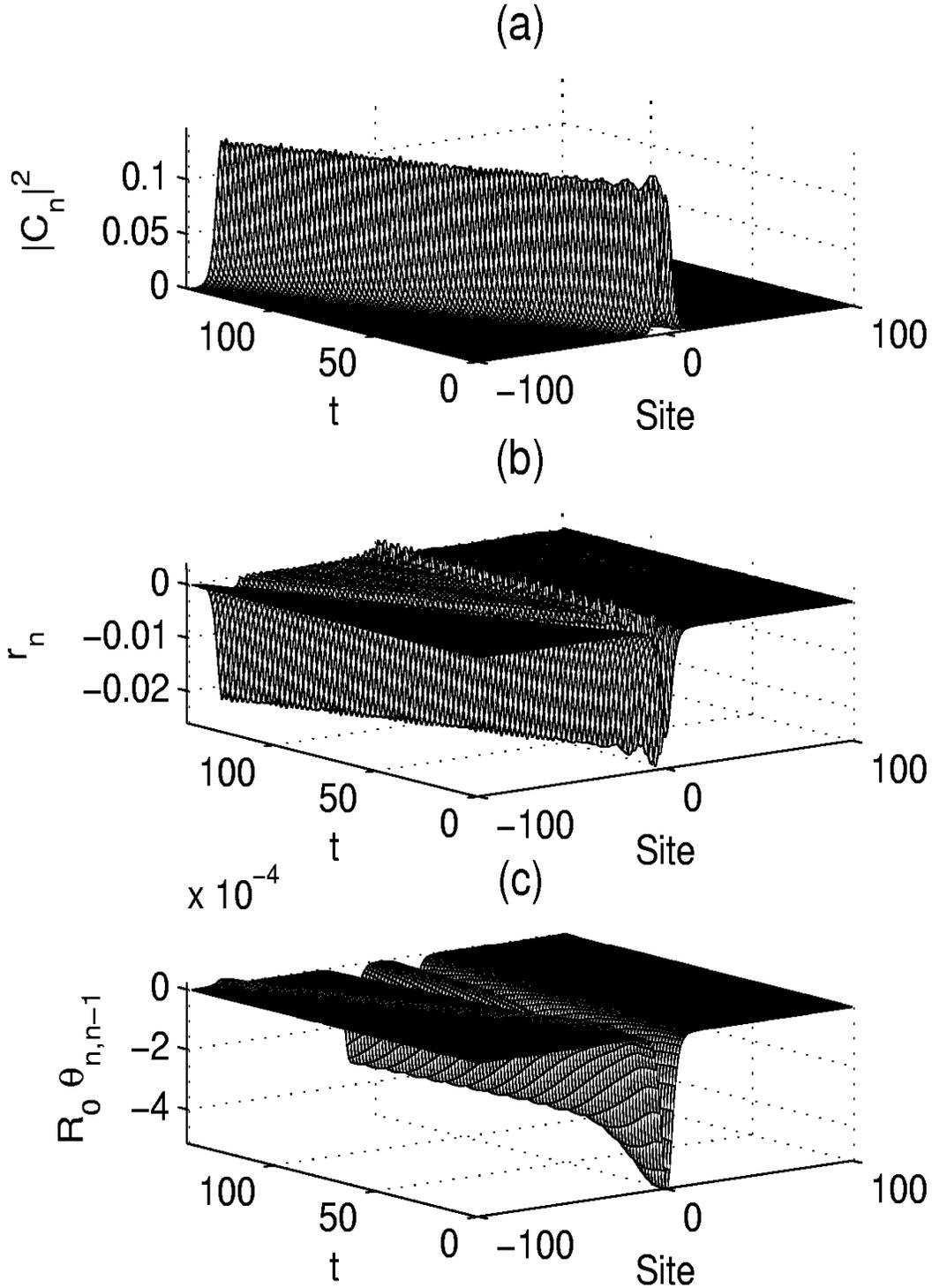}
  \end{center}
  \caption{Radial movability regime. Polaron motion along the chain in ordered case for
  $\alpha=0.0002$
   and an initial kick $\lambda_r=0.01$ in the radial velocities, corresponding to a
   kinetic energy of the perturbation $\Delta K=5 \times 10^{-6}$ eV.
  (a) Uniformly moving electronic excitation. (b) Radial displacements ($\AA$).
  (c) Arc displacements $R_0\,\theta_{n,n-1}$ ($\AA$) with $R_0=10\AA$.}
  \label{stationary_1}
\end{figure}

In absence of diagonal disorder, and if the parameter $\alpha$ is
small enough ($\alpha\lesssim 0.005$), it is not possible to move
the polarons by perturbing the angular variables. Mobility can only be
accomplished through perturbation of the radial ones, as in Ref.~\cite{HAA03}.
 This regime is explained in
detail in this reference, but, for completeness, we summarize here
the most relevant
characteristics:
\begin{enumerate}
\item The charge probability moves with uniform
velocity along the lattice with apparently constant profile.
\item The movement of the charge is accompanied by a relatively large radial
deformation ($r_n \sim 0.02 \, \AA$) and a negligible angular
lattice deformation ($R_0\theta_{n,n-1} \sim 2\cdot 10^{-5}\AA$).
\item The velocity of the polaron increases approximately
in a linear way with the kinetic
energy of the perturbation, as shown in Fig.~\ref{ve_o}, until a
critical value, beyond which the localization is destroyed.
\end{enumerate}
If we observe the spatial variables, there exists a small radial
amplitude oscillation pinned at the starting site. There are
radial and angular oscillations accompanying  the charge and two
larger angular oscillations, emerging from the same origin, that
propagate in opposite directions with lower velocity than the
polaron, as shown in Fig.~\ref{stationary_1}.

\subsection{Twist movability regime}
However, for larger values of parameter $\alpha$ ($\alpha\gtrsim
0.01$), the polaron can only be moved by perturbing the angular
variables. Perturbations of the radial variables cannot activate
mobility. This is a consequence of the different contributions of
the radial and angular components in the discrete gradient.

\begin{center}
\begin{figure}
\begin{center}
 \includegraphics[height=0.85\textheight,width=\textwidth]{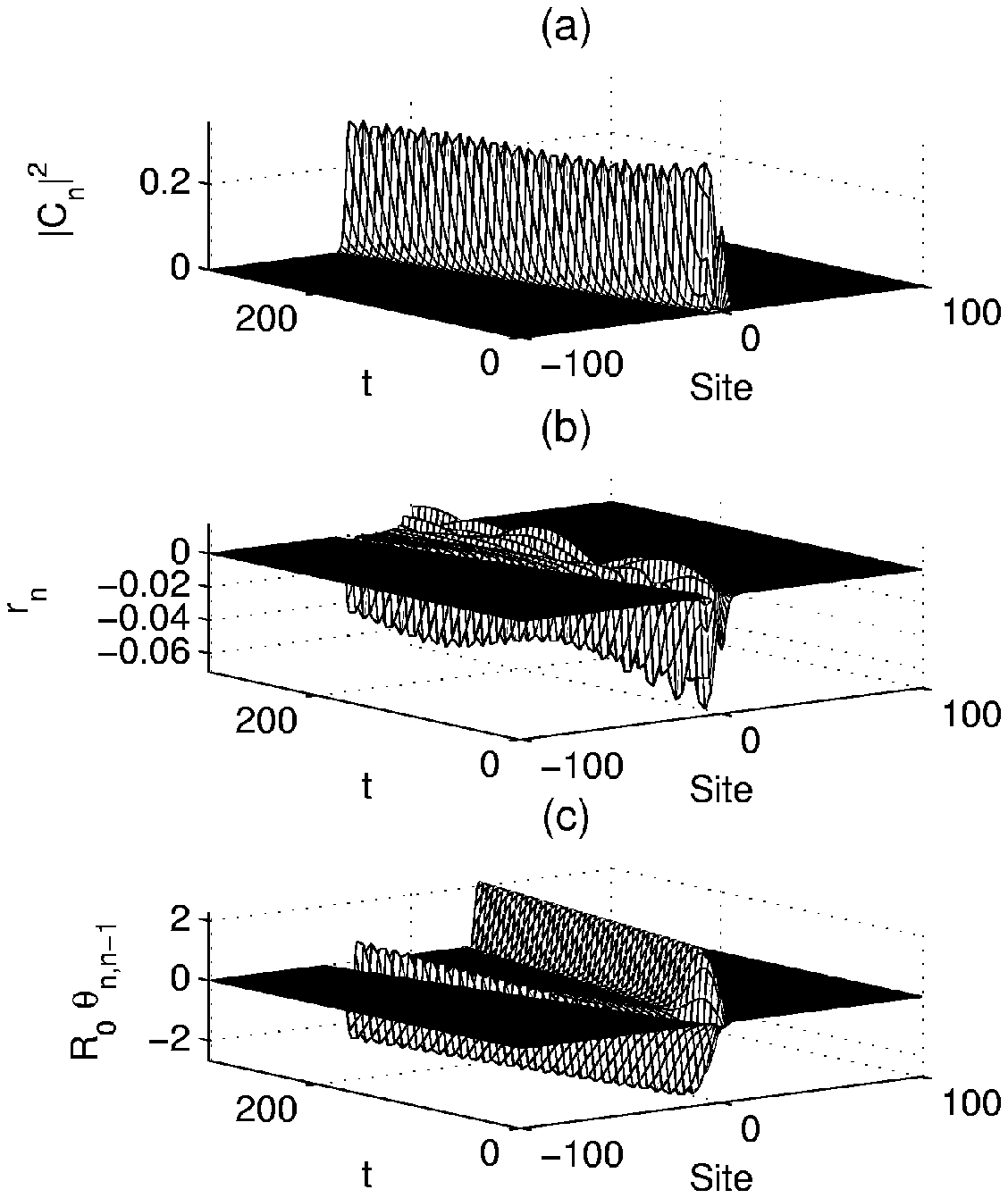}
  \caption{Twist movability regime. Polaron motion along the chain in ordered case for
  $\alpha=0.05$ and an initial kick $\lambda_{\varphi}=0.01$ in the angular
   velocities, corresponding to a
   kinetic energy of the perturbation $\Delta K=0.02$ eV.
  (a) Uniformly moving electronic excitation. (b) Radial displacements ($\AA$).
  (c) Arc displacements $R_0\theta_{n,n-1}$($\AA$) with $R_0=10\AA$.}
  \label{stationary_2}
\end{center}
\end{figure}
\end{center}

\begin{figure}
  \begin{center}
    \includegraphics[width=\singlefig]{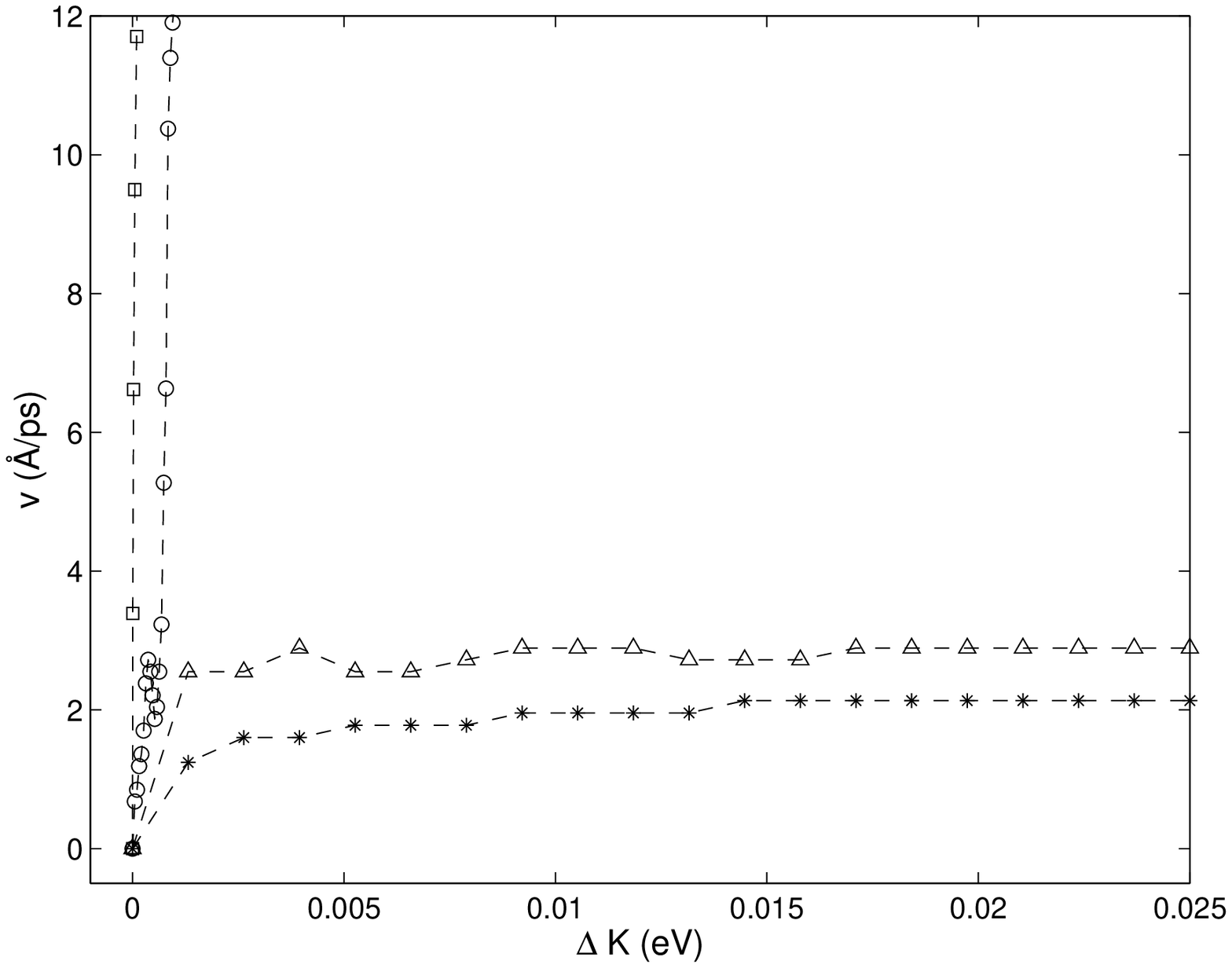}
  \end{center}
  \caption{Velocity of the polaron as a function of the
  kinetic energy of the perturbation $\Delta K$ added to the ground
  state   in the ordered case. Squares correspond to the
  radial movability regime ($\alpha=0.0002$). Circles correspond
  to the mixed regime with radial activation
  ($\alpha=0.01$). Triangles
  correspond to the mixed regime with angular activation for the same value of $\alpha=0.01$.
  Stars correspond to the
  twist movability regime ($\alpha=0.05$).}
  \label{ve_o}
\end{figure}

On the twist movability regime we observe that:
\begin{enumerate}
\item The charge probability apparently moves with
constant velocity and constant profile.
\item There are relatively large radial and angular deformations
travelling with the charge
 ( $r_n\sim 0.05 \,\AA$, $R_0\theta_{n,n-1}
\sim 2\,\AA$ ).
\item The velocity of the polaron increases initially with the
kinetic energy of the perturbation but tends to be approximately
constant, as  Fig.~\ref{ve_o} shows. It decreases slowly with
$\alpha$. This moving excitation remains localized even for large
perturbations.
\end{enumerate}
There remains a small radial oscillation localized at the starting
site. Also an angular deformation travels with opposite
direction and phase than the angular pulse moving with the charge.
\subsection{Mixed regime}
It is possible, for intermediates values of the parameter $\alpha$
($0.005\lesssim \alpha \lesssim 0.01$), that the polaron can
become mobile by perturbing any set of variables, the radial or
the angular ones. Nevertheless, the movement is rather different
as shown in Fig~\ref{ve_o}, when the polaron propagation is
activated by means of  radial perturbations, its characteristics
are similar to the radial movability regimen, and likewise when it
is activated by angular perturbations.

In general, radial movability requires less energy and has higher
velocity than the angular one, as shown in Fig.~\ref{ve_o}. There
exists a small interval of perturbations energies, for which the
movement can  be activated by means of both kind of perturbations.
The velocities obtained are similar but the evolution of the
spatial variables are rather different.

It is worth remarking that the limits of these regimes are not
exact, depending on the kinetic energy of the perturbation, but we
have intended to present a general picture.

\subsection{Tail analysis of the twist movability regime}
The consequences for the
tail analysis of the radial movability regime has been analyzed in Ref.~\cite{AHA03}.
Here we use this
simple but useful instrument to perform a similar study of the
twist movability regime. The tails of the moving polaron have
small amplitude and therefore their dynamics is to a great extent
linear. The linear  equations corresponding to
Eqs.~(\ref{electron}-\ref{angular}) in the ordered case are:
\begin{eqnarray}
\ii \tau \dot{c}_n =-c_{n+1}-c_{n-1}\,;\; \ddot{r}_n&=&-r_n
\,;\;
\ddot{\phi}_n=-\Omega^2\,(2\,\phi_n-\phi_{n-1}-\phi_{n+1}).
\label{eq:linear}
\end{eqnarray}
According to these equations, the variables $r_n$ are driven by
the nonlinear terms and we will not consider them here. We propose
the moving tail modes $c_n=\exp(-\xi/2\,(n-v\,t)-\ii\,E\,t/\tau)$
and $\phi_n=\exp(-\xi\,(n-v\,t))$, which will be valid
(relatively) far from the head of the moving polaron and while
$(n-v\,t)>0$. These modes comply with the observed phenomena of
similar exponential decay for $|c_n|^2$ and $\phi_n$ and that the
second magnitude is a deformation and not an oscillation.
Substitution into the first and third equations leads to:
\begin{eqnarray}
\tau\,\xi\,v=4\,\sin(q)\,\sinh(\xi/2),&\mbox{}& E=-2\,\cos(q)\,\cosh(\xi/2),\label{eq:linpol}\\
\; v&=&2\Omega\,\sinh(\xi/2)\,\xi^{-1} \label{eq:linv}.
\end{eqnarray}
The velocity in physical units  is $\tilde{v}=a\,\Omega_r\,v$. The
inverse characteristic length can be related to the participation
number by
$$P=\frac{(\sum_{n=-\infty}^{n=\infty}\exp(-\xi\,|n|))^2}
 {\sum_{n=-\infty}^{n=\infty}(\exp(-\xi\,|n|))^2}
=\frac{\tanh(\xi)}{\tanh^2(\xi/2)}\,.$$ Certainly $\xi$ and $P$
are determined by the full linear equations and depend on
$\alpha$, but in this way we can relate the values of $\xi$ with
the numerical results for $P$. Within the twist movability regime
the values of $P$ change from 5 to 12, which correspond to values
for $\xi$, 0.74 and 0.33, and for $v$, 3.17 and 3.11
($\AA/\mathrm{ps}$), respectively. For $\xi$ small, $P\sim 4/\xi$, $v\sim \Omega$
and $\tilde{v}=a\,\Omega_r\, \Omega=3.4\, \AA\,
8\,\mathrm{ps}^{-1}\sqrt{0.013}\sim 3.1\, \AA \,\mathrm{ps}^{-1}
$. The numerical results vary from 2.1 to 3.4, i.e., they are in
good agreement with the values of this tail analysis, taking into
account the drastic simplification of the model. Moreover, they
are also coherent with the numerical observation that the velocity of
the polarons practically does not depend on the kinetic energy of
the perturbation, as Fig.~\ref{ve_o} shows.  The conclusion is
that in this regime the polaron is basically {\em riding} on the
angular wave. Equations (\ref{eq:linpol}) can be used to obtain
the wave number $q$ and the charge energy $E$ and to compare them with
the numerical data with good results, with $q\sim 6\cdot 10^{-6}$
and energies slightly below $E_0=-2$ (scaled units). This last value corresponds to
the extended ground state with $q=0$ and $\xi=0$~\cite{AHA03}.

\section{Charge transport in the presence of disorder}

If an amount of  disorder in the on--site energies $E_n^0$ is
introduced, with random values $|E_n^0|<\Delta E$, we find that
moving polarons exist below a critical value  $\Delta E_{crit}$.
Beyond this value polarons cannot be moved as found in
Ref.~\cite{HAA03}. The most relevant results are:
\begin{enumerate}
\item In the radial movability regime, mobility is very sensitive to disorder. Only a
small degree of disorder allows  activation of polaron motion
(e.g., if $\alpha=0.0002$, $\Delta E_{crit}\approx 0.005$ eV).
\item In the twist movability regime the moving polaron is very
robust with respect to disorder (e.g., if $\alpha=0.05$, $\Delta
E_{crit}\approx 0.05$ eV).
\item For the mixed regime, angular activation leads to more robust (but slower)
polarons than radial activation.
If
$\Delta E$ is high enough, only angular activation is possible.
The movement is very similar to the ordered case.
\end{enumerate}

\section{Conclusions}

We have performed an analysis of the influence of the radial and
angular perturbations on the properties of moving polarons in a
three--dimensional, semi--classical, tight--binding model for DNA
in both the ordered and disordered case. There exist three different
regimes depending on the parameter $\alpha$ that describes the
coupling of the transfer integrals with the deformations of the
hydrogen bonds: a)~Radial movability, with charge transport
activated only by perturbing the radial variables; b)~Twist
movability, with charge transport activated by angular
perturbations; c)~Mixed regime.

The properties of the moving polarons are different in each regime.
In general, the mobility induced by angular activation  is more
robust with respect to  parametric disorder, the polarons have lower
velocity, and the activation energies are higher than in the radial
movability regime.

We have found that charge transport along the
DNA chain is strongly influenced by the value of $\alpha$. Therefore,
experimental observations of charge transport in real DNA could
help to determine this crucial magnitude, which, in turn, is
necessary to understand the phenomenon of charge transport in DNA.

\section*{Acknowledgments}
 The authors are grateful to partial support under the LOCNET EU network
HPRN-CT-1999-00163. JFRA  acknowledges DH and the
 Institut f\"{u}r Theoretische Physik for their warm hospitality

\newcommand{\noopsort}[1]{} \newcommand{\printfirst}[2]{#1}
  \newcommand{\singleletter}[1]{#1} \newcommand{\switchargs}[2]{#2#1}

\end{document}